%% file: main.tex
\documentclass[sigconf,nonacm]{acmart}

\usepackage{booktabs}
\usepackage{multirow}
\usepackage{array}
\usepackage{float}
\usepackage{marvosym}
\newcolumntype{L}[1]{>{\raggedright\arraybackslash}p{#1}}
\newcolumntype{C}[1]{>{\centering\arraybackslash}p{#1}}

\makeatletter
\renewcommand{\country}[1]{\global\@ACM@countrypresenttrue}
\makeatother

\begin{document}
\title{Numerical Embeddings for CTR Prediction: A Taxonomy, Unified Benchmark, and Open-Source FuxiCTR Extension}

\author{Heng Yao\textsuperscript{*\,\Letter}}
\affiliation{\institution{Ant Group}\country{}}

\author{Hanwei Zhu\textsuperscript{*}}
\affiliation{\institution{Independent Researcher}\country{}}

\author{Tianying Liu}
\affiliation{\institution{Independent Researcher}\country{}}

\author{Yulou Shu}
\affiliation{\institution{Alibaba Inc.}\country{}}

\author{Siyun Hou\textsuperscript{\Letter}}
\affiliation{\institution{Henan Polytechnic University}\country{}}

\renewcommand{\shortauthors}{Yao et al.}

\begin{abstract}
Numerical embeddings map scalar features into vectors before a click-through rate (CTR) backbone combines the fields. To our knowledge, NumEmbBench is the first open benchmark dedicated to comparing numerical embeddings for CTR prediction. We classify 18 configurations into six mechanism-based families and evaluate them through one FuxiCTR interface. After equal-budget validation tuning, we compare frozen configurations on three datasets, nine backbones, and five seeds under a shared train-only z-score protocol. DEER has the best mean rank, but six methods lead at least one dataset--backbone setting. A matched native-scale test changes the winner in 11 of 27 settings, while paired latency measurements show that accuracy and inference cost need not move together. Code and per-seed results: \href{https://github.com/YH-learning/NumEmbBench}{\textcolor{blue}{\nolinkurl{https://github.com/YH-learning/NumEmbBench}}}.
\end{abstract}

\maketitle

\begingroup
\makeatletter
\def\@makefnmark{\hbox{\@textsuperscript{*}}}
\footnotetext{Heng Yao and Hanwei Zhu contributed equally.}
\def\@makefnmark{\hbox{\@textsuperscript{\Letter}}}
\footnotetext{Corresponding authors: Heng Yao and Siyun Hou. E-mail: \href{mailto:yaoheng.cs@qq.com}{yaoheng.cs@qq.com} (Heng Yao); \href{mailto:housiyun@hpu.edu.cn}{housiyun@hpu.edu.cn} (Siyun Hou).}
\makeatother
\endgroup

\section{Introduction}

CTR models combine categorical fields, such as user or item identifiers, with numerical fields, such as prices and counts. Categorical identifiers use learned lookup vectors. A numerical value instead has magnitude and order, so nearby values may carry related information. A \emph{numerical embedding} maps scalar $x_j$ to vector $\mathbf e_j=f_j(x_j)\in\mathbb R^d$. Thus the backbone sees a description of the value, such as its scale, interval, or smooth basis responses, rather than the raw number alone. Some methods also condition this vector on other fields~\cite{guo2021autodis,gorishniy2022numemb,cheng2022deer,chen2022narydis,liu2026daes}.

General tabular studies have compared numerical encoders across backbones~\cite{gorishniy2022numemb}, and CTR method papers include numerical-encoder baselines~\cite{guo2021autodis,liu2026daes}. BARS instead benchmarks CTR backbones~\cite{zhu2021bars}. These studies do not jointly provide an open, encoder-centered CTR benchmark. Their preprocessing, models, and training pipelines also differ, so an apparent encoder gain may reflect the surrounding setup. A dedicated comparison must isolate the numerical representation and measure accuracy and computational cost.

We ask which encoders remain competitive across backbones, whether their rankings transfer across input scales, and what training and inference cost accompanies their accuracy. A shared z-score protocol defines the primary leaderboard; a native-scale track tests the same frozen configurations without retuning.

\noindent Our contributions are threefold:
\begin{itemize}
\setlength{\itemsep}{0pt}\setlength{\topsep}{2pt}\setlength{\parsep}{0pt}
\item We introduce, to our knowledge, \textbf{the first open CTR benchmark centered on numerical embeddings} and organize 18 configurations into six mechanism-based families.
\item We establish an \textbf{equal-budget, five-seed comparison} across three datasets and nine CTR backbones, including input-scale and efficiency analyses.
\item We release a \textbf{reusable FuxiCTR plugin interface}, data-processing scripts, frozen configurations, and per-seed results for evaluating future encoders under the same protocol.
\end{itemize}

\section{Taxonomy and Benchmark Design}

\subsection{Numerical Embedding Taxonomy}

\begin{table*}[t]
\caption{Six families classified by their numerical coordinate construction. Statistics means a mapping uses training-derived numerical statistics; context means it can depend on fields other than the scalar being embedded.}
\label{tab:taxonomy}
\centering\small
\setlength{\tabcolsep}{2.5pt}
\begin{tabular*}{\textwidth}{@{\extracolsep{\fill}}L{3.4cm}L{5.1cm}L{5.3cm}cc@{}}
\toprule
Family & Evaluated configurations & Defining mechanism & Statistics & Context \\
\midrule
Pointwise projection & Field, Linear, Linear-ReLU & Scale or transform one scalar & No & No \\
Discretization & Quantile, LogBucket, AutoDis~\cite{guo2021autodis} & Assign intervals or learned prototypes & Optional & No \\
Fixed basis & Periodic, B-spline & Evaluate preset basis functions & No & No \\
Piecewise basis & PLE-A/B/ReLU, PLR~\cite{gorishniy2022numemb} & Construct local interval coordinates & Yes & No \\
Adaptive coordinates & NaryDis/Lite~\cite{chen2022narydis}, DEER~\cite{cheng2022deer} & Learn local geometry or symbolic digits & Optional & No \\
Distribution/context adaptive & DAE~\cite{shen2023dae}, DAES-Gate/Tran~\cite{liu2026daes} & Adapt to distributions or selected context & Yes & Method dependent \\
\bottomrule
\end{tabular*}
\end{table*}

Let $x_j\in\mathbb{R}$ be numerical field $j$ and $\mathbf e_j=f_j(x_j)\in\mathbb{R}^{16}$ its backbone input. Table~\ref{tab:taxonomy} classifies methods by the mechanism that constructs $\mathbf e_j$, not by the downstream interaction network. The plugin registry exposes the same field-wise interface to DNN, DeepFM, WideDeep, NFM, PNN, DCN, DCNv2, AutoInt, and FiBiNET~\cite{cheng2016widedeep,guo2017deepfm,he2017nfm,qu2016pnn,wang2017dcn,wang2021dcnv2,song2019autoint,huang2019fibinet}. Categorical embeddings and each backbone's native interaction structure remain intact.

\paragraph{Coordinate mechanisms.}
Pointwise methods use the scalar directly or apply a small learned projection. Quantile and logarithmic buckets turn it into a discrete region; AutoDis instead assigns differentiable weights to learned prototypes~\cite{guo2021autodis}. Periodic and spline bases evaluate multiple functions of the same input, while PLE and PLR construct continuous, local coordinates over intervals~\cite{gorishniy2022numemb}. DEER learns local numerical geometry and NaryDis uses a symbolic multi-radix representation~\cite{cheng2022deer,chen2022narydis}. DAE adapts to feature distributions, and DAES additionally uses selected contextual fields~\cite{shen2023dae,liu2026daes}. Grouping these methods by the source of their coordinates makes similarities and differences testable without treating architectural names as families.

The families differ along three representation axes. First, a hard bucket maps nearby values in the same bin to one code and can break continuity at a boundary, whereas soft assignments and basis expansions can preserve graded changes. Second, a preset basis evaluates a fixed set of functions across the input range, while piecewise methods use interval-local support and adaptive-coordinate methods construct their own geometry or digits. Third, most encoders depend only on the target scalar after fitting; distribution-adaptive methods use training-derived information, and selected DAES variants also condition on other fields. These distinctions describe the numerical information supplied to the backbone, not the architecture of the backbone itself.

\subsection{Evaluation Protocol and Reproducibility}

\paragraph{Common interface.}
Every plugin receives field-wise numerical tensors and returns one vector of the same width for each field. A registry selects the implementation, and method-specific adapters supply training-derived statistics or selected context only where required. Thus the numerical branch changes without rewriting each CTR backbone. The release distinguishes direct implementations, wrappers around official dependencies, and paper-based reimplementations; tests cover output shape, field ordering, finite gradients, and serialization. This is an infrastructure contribution, not a claim that the underlying CTR library is new~\cite{zhu2021bars}.

\paragraph{Data and shared settings.}
The three datasets are AutoML-A, AutoML-E, and Criteo, with 2.80M, 5.14M, and 36.67M training examples, respectively~\cite{guyon2019automl}. All numerical statistics are fitted on training data only. The primary protocol applies the same train-only z-score transform to every encoder. For each dataset--backbone setting (a cell), we use fixed splits, batch size 4096, Adam at $10^{-3}$, validation-AUC checkpointing, patience 2, and seeds 2026--2030.

The datasets have 23, 6, and 13 numerical fields, respectively, alongside 51, 25, and 26 categorical fields. Processing scripts record split manifests and train-derived vocabularies so that an encoder cannot infer validation or test statistics. All methods output 16-dimensional numerical vectors. Each backbone's linear and interaction paths are configured consistently across encoders; the exact numerical-field wiring is recorded with the implementation.

\paragraph{Tuning and evaluation.}
Each method receives the same staged tuning budget: a bounded screening round on DNN and DCNv2, followed by full-schedule validation of the two best candidates on those development backbones. One configuration per method is frozen by validation rank, with logloss breaking ties. The final nine-backbone evaluation contains $18\times3\times9\times5=2{,}430$ runs; test metrics never select configurations. A second 2,430-run matrix changes only the numerical input scale and its train-derived LogBucket vocabulary, reusing those frozen configurations and seeds. Complete per-seed matrices and preprocessing scripts are in the public code repository.

This design deliberately separates three questions: the primary leaderboard measures representation quality after a common stabilizing transform; the native-scale track measures whether the chosen configurations transfer to observed input ranges; and the efficiency study measures implementation cost under a common measurement protocol. In the tuning screen, candidates are predeclared for each encoder and receive the same evaluation budget. The final global configuration is not customized to each backbone or test split, avoiding a large hidden per-cell search advantage.

\section{Results and Implications}

\input{tables/paired_scale_auc_results}

\subsection{RQ1: Which Encoders Lead Across Backbones?}

We first compare the \emph{z-score} columns of Table~\ref{tab:full-results}, where every encoder receives the same train-only z-score preprocessing. Across the 27 dataset--backbone cells, DEER has the best mean AUC rank (3.41), 14 wins, and 19 top-three placements; PLE-B follows at 5.30. A Friedman test rejects equal method performance ($\chi^2=157.04$, $p=1.02\times10^{-24}$), and paired comparisons against DEER remain significant after Holm correction. This supports DEER as a strong default under the primary protocol, not as a universal winner.

The full table shows where the summary breaks down. DEER leads eight of nine Criteo backbones but only one of nine AutoML-E backbones; six methods lead at least one z-score cell overall. Even on AutoML-A, the DEER--Periodic ordering reverses: DeepFM yields 0.7103 versus 0.6982, while FiBiNET yields 0.7063 versus 0.7110. These are five-seed means, not claims that either pairwise difference is statistically significant. A single-backbone leaderboard would miss such changes.

The taxonomy explains representation mechanisms, not a family-level performance ordering. Within piecewise basis, PLE-B ranks 5.30 on average but PLE-ReLU ranks 12.48; within adaptive coordinates, DEER ranks 3.41 but NaryDis-Lite ranks 13.44. Even Field leads three z-score cells. Thus the particular encoder and backbone matter more than a family label. We retain every dataset--backbone result instead of replacing the matrix with one aggregate score. The \emph{Raw} columns address scale transfer; per-seed AUC and logloss are available in the artifact.

\subsection{RQ2: Do Rankings Transfer to Raw Input Scale?}

Each \emph{Raw} column in Table~\ref{tab:full-results} is paired with the adjacent \emph{z-score} column: the split, seed, backbone, and configuration selected under z-score preprocessing remain fixed while the numerical input scale changes (with a train-derived LogBucket vocabulary). The winning method changes in 11 of 27 cells: one on Criteo, three on AutoML-A, and seven on AutoML-E. Scale transfer is therefore dataset dependent, rather than a uniform shift in AUC.\footnote{The mean within-cell rank correlation is 0.369; complete per-seed AUC and logloss for both scales are in the public code repository.}

The differences also vary sharply by encoder within the same backbone. On AutoML-E/AutoInt, PLE-B changes from 0.8411 to 0.5000, whereas B-spline stays near its z-score result (0.8417 versus 0.8420). Across the matrix, 57 of 486 Raw method cells have test AUC below 0.55 (14 on AutoML-A and 43 on AutoML-E, none on Criteo), compared with none under z-score preprocessing. The concentration on AutoML-E cautions against generalizing a single dataset's apparent scale robustness. This tests transfer, not independently tuned Raw performance; retuning may recover low scores. Yet the same backbone and training budget do not ensure scale robustness, so preprocessing must accompany each encoder comparison.

\subsection{RQ3: What Does the Accuracy Cost?}

\begin{table}[H]
\caption{Numerical parameters and paired training and inference latency relative to Field. Ratios use the same dataset, backbone, and measurement round.}
\label{tab:efficiency}
\centering
\footnotesize
\setlength{\tabcolsep}{1.5pt}
\begin{tabular*}{\columnwidth}{@{\extracolsep{\fill}}llrrr@{}}
\toprule
Family & Method & Param. & Train & Infer. \\
\midrule
\multirow{3}{*}{\shortstack[l]{Pointwise\\projection}} & Field & 0.2k & 1.00$\times$ & 1.00$\times$ \\
 & Linear & 0.4k & 1.00$\times$ & 1.01$\times$ \\
 & Linear-ReLU & 0.4k & 1.01$\times$ & 1.02$\times$ \\
\midrule
\multirow{3}{*}{Discretization} & Quantile & 2.1k & 1.22$\times$ & 1.23$\times$ \\
 & LogBucket & 3.0k & 1.32$\times$ & 1.60$\times$ \\
 & AutoDis & 7.4k & 1.04$\times$ & 1.12$\times$ \\
\midrule
\multirow{2}{*}{Fixed basis} & Periodic & 3.9k & 1.03$\times$ & 1.07$\times$ \\
 & B-spline & 7.2k & 1.07$\times$ & 1.37$\times$ \\
\midrule
\multirow{4}{*}{Piecewise basis} & PLE-A & 3.8k & 1.02$\times$ & 1.09$\times$ \\
 & PLE-B & 4.0k & 1.03$\times$ & 1.10$\times$ \\
 & PLE-ReLU & 4.0k & 1.03$\times$ & 1.11$\times$ \\
 & PLR & 7.6k & 1.03$\times$ & 1.08$\times$ \\
\midrule
\multirow{3}{*}{\shortstack[l]{Adaptive\\coordinates}} & NaryDis-Lite & 19.0k & 1.25$\times$ & 2.26$\times$ \\
 & NaryDis & 32.3k & 1.44$\times$ & 2.43$\times$ \\
 & DEER & 4.0k & 0.76$\times$ & 1.20$\times$ \\
\midrule
\multirow{3}{*}{\shortstack[l]{Distribution/\\context adaptive}} & DAE & 36.7k & 1.08$\times$ & 1.13$\times$ \\
 & DAES-Gate & 4.5k & 1.28$\times$ & 1.31$\times$ \\
 & DAES-Tran & 60.9k & 1.32$\times$ & 1.34$\times$ \\
\bottomrule
\end{tabular*}
\end{table}

Table~\ref{tab:efficiency} adds cost to the z-score accuracy comparison in Table~\ref{tab:full-results}; we did not measure a separate Raw-scale latency matrix. We time real resident batches on the same idle GPU in eager FP32, using three representative backbones and three randomized rounds per method--dataset--backbone case; disk input is excluded. The 486 successful measurements use the same frozen configurations as the z-score accuracy study. DEER, the best method by mean z-score rank, uses about 4.0k numerical parameters and takes 0.76$\times$ Field training time but 1.20$\times$ Field inference time. The evaluated NaryDis variants exceed 2$\times$ Field inference latency. These ratios describe released implementations, not theoretical lower bounds.

Parameter count alone is an inadequate cost proxy. DAE has 36.7k numerical parameters yet 1.13$\times$ Field inference latency, whereas NaryDis-Lite has 19.0k parameters and 2.26$\times$ latency. PLE-B, second by mean z-score rank, takes 1.10$\times$ Field inference time compared with DEER's 1.20$\times$; its accuracy--latency tradeoff therefore merits examination in individual cells. Periodic, PLE-A/B, PLR, AutoDis, and DAE each remain within 1.13$\times$ Field at inference, but other discretization and adaptive-coordinate implementations are substantially slower. The family label identifies a representation mechanism, not a fixed computational price. Absolute latency, embedding-stage timing, and peak-memory records in the artifact support deployment-specific decisions; these ratios are not hardware-independent limits.

\paragraph{Overall takeaways.}
\begin{enumerate}
\setlength{\itemsep}{0pt}\setlength{\topsep}{1pt}\setlength{\parsep}{0pt}
\item \textbf{There is no universal accuracy winner.} DEER leads the shared z-score matrix by mean rank, yet six methods win at least one of its 27 dataset--backbone cells; even variants in one family differ substantially.
\item \textbf{Input scale changes model selection.} With configurations held fixed, the winner changes in 11 of 27 paired cells and 57 Raw method cells fall below 0.55 AUC, versus none under z-score preprocessing. These are transfer failures, not proof that Raw-specific retuning cannot help.
\item \textbf{Accuracy does not determine computational cost.} DEER trains at 0.76$\times$ Field time but infers at 1.20$\times$; the evaluated NaryDis variants exceed 2$\times$ inference time. Both training and serving measurements belong beside AUC.
\end{enumerate}
Together, these findings call for reporting the dataset, backbone, preprocessing, per-cell AUC, and paired cost rather than a single aggregate leaderboard.

\section{Open-Source Benchmark Artifact}

The release contains dataset converters, train-only preprocessing, encoder implementations, frozen configurations, a common runner, and deterministic table builders. Per-seed tables identify the dataset, backbone, encoder, seed, and validation and test AUC and logloss. Task manifests and configuration snapshots document the run settings, while the builders check completeness before aggregation. Provenance notes distinguish official dependencies, wrappers, and paper-based reimplementations. Thus the summary tables remain auditable without distributing model checkpoints.

The same artifact supports extension: a new encoder declares whether it needs fitted numerical statistics or other-field context and returns the common embedding width in the recorded field order. It can then enter the existing dataset--backbone--seed matrix without modifying the nine CTR backbones. This makes the release a reusable evaluation substrate, not only a snapshot of the present leaderboard.

\section{Conclusion and Limitations}

NumEmbBench turns numerical representation into a controlled model-selection problem: the taxonomy identifies distinct mechanisms, the common plugin interface separates them from CTR backbones, and the multi-backbone results expose ranking changes hidden by isolated comparisons. DEER is the strongest default under the shared z-score protocol, but its advantage does not extend uniformly to other members of its family or to every backbone. The benchmark is limited to three datasets, one global tuning budget per method, and the released eager implementations. AutoML-A and AutoML-E also share a source family, so the three datasets are not three independent domains. Native-scale results transfer configurations selected under z-score preprocessing rather than retune each encoder. Latency reflects the released implementations, not a hardware-independent limit.

\clearpage
\section*{Ethical Considerations}
The benchmark uses existing tabular CTR datasets and aggregate predictive metrics, with no new user study or personal-data collection. Better CTR prediction may reinforce exposure and feedback biases; deployment should separately audit privacy, fairness, and drift.

\bibliographystyle{ACM-Reference-Format}
\bibliography{references}
\end{document}

%% file: tables/paired_scale_auc_results.tex
\begin{table*}[t]
\caption{Five-seed mean test AUC under shared train-only z-score preprocessing and original numerical input scale (Raw). The Raw track reuses the configurations selected under z-score preprocessing without retuning. Bold marks the best AUC within each dataset--backbone--scale column. Full per-seed AUC and logloss are in the public code repository.}
\label{tab:full-results}
\centering
\renewcommand{\arraystretch}{0.68}
\setlength{\tabcolsep}{1.85pt}
\resizebox{0.94\textwidth}{!}{%
\begin{tabular}{llrrrrrrrrrrrrrrrrrr}
\toprule
\multicolumn{20}{c}{\textbf{Criteo}} \\
\midrule
Family & Method & \multicolumn{2}{c}{DNN} & \multicolumn{2}{c}{DeepFM} & \multicolumn{2}{c}{WideDeep} & \multicolumn{2}{c}{NFM} & \multicolumn{2}{c}{PNN} & \multicolumn{2}{c}{DCN} & \multicolumn{2}{c}{DCNv2} & \multicolumn{2}{c}{AutoInt} & \multicolumn{2}{c}{FiBiNET} \\
\cmidrule(lr){3-4}\cmidrule(lr){5-6}\cmidrule(lr){7-8}\cmidrule(lr){9-10}\cmidrule(lr){11-12}\cmidrule(lr){13-14}\cmidrule(lr){15-16}\cmidrule(lr){17-18}\cmidrule(lr){19-20}
 &  & z-score & Raw & z-score & Raw & z-score & Raw & z-score & Raw & z-score & Raw & z-score & Raw & z-score & Raw & z-score & Raw & z-score & Raw \\
\midrule
\multirow{3}{*}{\shortstack[l]{Pointwise\\projection}} & Field & 0.8085 & 0.8060 & 0.8056 & 0.7935 & 0.8079 & 0.8029 & 0.8009 & 0.7975 & 0.8092 & 0.8078 & 0.8083 & 0.8029 & 0.8094 & 0.8039 & 0.8089 & 0.8056 & 0.8090 & 0.8048 \\
 & Linear & 0.8086 & 0.8052 & 0.8041 & 0.7906 & 0.8077 & 0.7867 & 0.8010 & 0.7812 & 0.8091 & 0.8013 & 0.8082 & 0.7999 & 0.8094 & 0.8027 & 0.8088 & 0.8029 & 0.8095 & 0.7937 \\
 & Linear-ReLU & 0.8093 & 0.8080 & 0.8053 & 0.8023 & 0.8085 & 0.7968 & 0.8022 & 0.7817 & 0.8095 & 0.8098 & 0.8092 & 0.8078 & 0.8108 & 0.8108 & 0.8092 & 0.8050 & 0.8102 & 0.7960 \\
\midrule
\multirow{3}{*}{\shortstack[l]{Discretization}} & Quantile & 0.8101 & 0.8101 & 0.8076 & 0.8076 & 0.8093 & 0.8093 & 0.8052 & 0.8052 & 0.8112 & 0.8112 & 0.8101 & 0.8101 & 0.8116 & 0.8116 & 0.8105 & 0.8105 & 0.8108 & 0.8108 \\
 & LogBucket & 0.7936 & 0.8097 & 0.7913 & 0.8069 & 0.7930 & 0.8090 & 0.7900 & 0.8044 & 0.7945 & 0.8108 & 0.7937 & 0.8097 & 0.7947 & 0.8113 & 0.7939 & 0.8099 & 0.7937 & 0.8103 \\
 & AutoDis & 0.8102 & 0.8055 & 0.8077 & 0.8027 & 0.8095 & 0.8045 & 0.8043 & 0.8007 & 0.8110 & 0.8062 & 0.8103 & 0.8052 & 0.8115 & 0.8068 & 0.8105 & 0.8055 & 0.8104 & 0.8059 \\
\midrule
\multirow{2}{*}{\shortstack[l]{Fixed basis}} & Periodic & 0.8090 & 0.8057 & 0.8023 & 0.8023 & 0.8080 & 0.8046 & 0.8028 & 0.8000 & 0.8097 & 0.8070 & 0.8088 & 0.8055 & 0.8107 & 0.8083 & 0.8086 & 0.8057 & 0.8100 & 0.8069 \\
 & B-spline & 0.8092 & 0.8092 & 0.8065 & 0.8064 & 0.8085 & 0.8086 & 0.8013 & 0.8013 & 0.8095 & 0.8095 & 0.8093 & 0.8093 & 0.8104 & 0.8104 & 0.8094 & 0.8093 & 0.8097 & 0.8097 \\
\midrule
\multirow{4}{*}{\shortstack[l]{Piecewise basis}} & PLE-A & 0.8104 & 0.8104 & 0.8065 & 0.8065 & 0.8096 & 0.8096 & 0.8049 & 0.8049 & 0.8115 & 0.8114 & 0.8103 & 0.8103 & 0.8122 & 0.8122 & 0.8105 & 0.8105 & 0.8116 & \textbf{0.8116} \\
 & PLE-B & 0.8106 & 0.8065 & 0.8065 & 0.7973 & 0.8100 & 0.7939 & 0.8052 & 0.7772 & 0.8118 & 0.8024 & 0.8107 & 0.8050 & 0.8125 & 0.8055 & 0.8108 & 0.8040 & \textbf{0.8117} & 0.7884 \\
 & PLE-ReLU & 0.8085 & 0.8052 & 0.8040 & 0.7924 & 0.8079 & 0.7768 & 0.8008 & 0.7781 & 0.8091 & 0.8016 & 0.8083 & 0.8017 & 0.8096 & 0.8034 & 0.8088 & 0.8032 & 0.8095 & 0.7837 \\
 & PLR & 0.8095 & 0.8067 & 0.8064 & 0.8033 & 0.8090 & 0.8062 & 0.8037 & 0.8010 & 0.8103 & 0.8075 & 0.8096 & 0.8068 & 0.8113 & 0.8085 & 0.8099 & 0.8067 & 0.8103 & 0.8078 \\
\midrule
\multirow{3}{*}{\shortstack[l]{Adaptive\\coordinates}} & NaryDis-Lite & 0.8093 & 0.8106 & 0.8066 & 0.8081 & 0.8085 & 0.8098 & 0.8017 & 0.8047 & 0.8099 & 0.8116 & 0.8092 & 0.8105 & 0.8110 & 0.8121 & 0.8096 & 0.8108 & 0.8096 & 0.8111 \\
 & NaryDis & 0.8104 & 0.8108 & 0.8080 & 0.8083 & 0.8097 & 0.8101 & 0.8053 & 0.8059 & 0.8112 & 0.8118 & 0.8104 & 0.8107 & 0.8119 & 0.8123 & 0.8106 & 0.8111 & 0.8110 & 0.8114 \\
 & DEER & \textbf{0.8109} & \textbf{0.8110} & \textbf{0.8085} & \textbf{0.8086} & \textbf{0.8103} & \textbf{0.8103} & \textbf{0.8059} & \textbf{0.8059} & \textbf{0.8122} & \textbf{0.8122} & \textbf{0.8110} & \textbf{0.8109} & \textbf{0.8126} & \textbf{0.8126} & \textbf{0.8114} & \textbf{0.8114} & 0.8116 & 0.8115 \\
\midrule
\multirow{3}{*}{\shortstack[l]{Distribution/\\context adaptive}} & DAE & 0.8103 & 0.7984 & 0.8078 & 0.7948 & 0.8095 & 0.7975 & 0.8050 & 0.7942 & 0.8108 & 0.7973 & 0.8103 & 0.7993 & 0.8116 & 0.7969 & 0.8104 & 0.8006 & 0.8104 & 0.7982 \\
 & DAES-Gate & 0.8106 & 0.8106 & 0.8076 & 0.8075 & 0.8098 & 0.8098 & 0.8057 & 0.8058 & 0.8117 & 0.8117 & 0.8105 & 0.8106 & 0.8122 & 0.8123 & 0.8107 & 0.8107 & 0.8114 & 0.8114 \\
 & DAES-Tran & 0.8105 & 0.8104 & 0.8070 & 0.8072 & 0.8097 & 0.8097 & 0.8058 & 0.8058 & 0.8111 & 0.8110 & 0.8102 & 0.8102 & 0.8121 & 0.8120 & 0.8100 & 0.8101 & 0.8113 & 0.8113 \\
\midrule
\addlinespace[2pt]
\multicolumn{20}{c}{\textbf{AutoML-A}} \\
\midrule
Family & Method & \multicolumn{2}{c}{DNN} & \multicolumn{2}{c}{DeepFM} & \multicolumn{2}{c}{WideDeep} & \multicolumn{2}{c}{NFM} & \multicolumn{2}{c}{PNN} & \multicolumn{2}{c}{DCN} & \multicolumn{2}{c}{DCNv2} & \multicolumn{2}{c}{AutoInt} & \multicolumn{2}{c}{FiBiNET} \\
\cmidrule(lr){3-4}\cmidrule(lr){5-6}\cmidrule(lr){7-8}\cmidrule(lr){9-10}\cmidrule(lr){11-12}\cmidrule(lr){13-14}\cmidrule(lr){15-16}\cmidrule(lr){17-18}\cmidrule(lr){19-20}
 &  & z-score & Raw & z-score & Raw & z-score & Raw & z-score & Raw & z-score & Raw & z-score & Raw & z-score & Raw & z-score & Raw & z-score & Raw \\
\midrule
\multirow{3}{*}{\shortstack[l]{Pointwise\\projection}} & Field & 0.7054 & 0.6648 & 0.7074 & 0.5728 & 0.7067 & 0.6585 & 0.6857 & 0.6680 & 0.7011 & 0.6553 & 0.7063 & 0.6596 & 0.7078 & 0.6233 & 0.7111 & 0.6497 & 0.6979 & 0.6284 \\
 & Linear & 0.7052 & 0.6157 & 0.6966 & 0.4949 & 0.7045 & 0.5650 & 0.6881 & 0.5342 & 0.7051 & 0.5706 & 0.7051 & 0.5028 & 0.7056 & 0.5264 & 0.7064 & 0.6508 & 0.7097 & 0.5677 \\
 & Linear-ReLU & 0.7044 & 0.6418 & 0.6205 & 0.5000 & 0.7031 & 0.5845 & 0.6812 & 0.5687 & 0.7040 & 0.6732 & 0.7037 & 0.5410 & 0.7047 & 0.6813 & 0.7081 & 0.6440 & 0.7080 & 0.5598 \\
\midrule
\multirow{3}{*}{\shortstack[l]{Discretization}} & Quantile & 0.7058 & 0.7058 & 0.7078 & 0.7082 & 0.7067 & 0.7072 & 0.6910 & 0.6912 & 0.7054 & 0.7052 & 0.7045 & 0.7047 & 0.7059 & 0.7061 & 0.7094 & 0.7099 & 0.7046 & 0.7043 \\
 & LogBucket & 0.6804 & 0.6822 & 0.6798 & 0.6847 & 0.6790 & 0.6823 & 0.6645 & 0.6691 & 0.6799 & 0.6820 & 0.6791 & 0.6792 & 0.6805 & 0.6803 & 0.6862 & 0.6882 & 0.6798 & 0.6812 \\
 & AutoDis & 0.7061 & 0.6895 & 0.7065 & 0.6857 & 0.7067 & 0.6847 & 0.6887 & 0.6705 & 0.7065 & 0.6892 & 0.7039 & 0.6889 & 0.7078 & 0.6870 & 0.7132 & 0.7075 & 0.7047 & 0.6891 \\
\midrule
\multirow{2}{*}{\shortstack[l]{Fixed basis}} & Periodic & 0.7050 & 0.6847 & 0.6982 & 0.6655 & 0.7024 & 0.6850 & 0.6921 & 0.6854 & 0.7038 & 0.6841 & 0.7025 & 0.6814 & 0.7084 & 0.6867 & 0.7086 & 0.6834 & \textbf{0.7110} & 0.6814 \\
 & B-spline & 0.7079 & 0.7075 & 0.7096 & 0.7097 & 0.7087 & 0.7088 & 0.6926 & \textbf{0.6927} & 0.7060 & 0.7060 & 0.7068 & 0.7066 & 0.7090 & \textbf{0.7088} & 0.7133 & 0.7137 & 0.7100 & 0.7090 \\
\midrule
\multirow{4}{*}{\shortstack[l]{Piecewise basis}} & PLE-A & 0.7078 & 0.7075 & 0.6761 & 0.6762 & 0.7011 & 0.7014 & 0.6898 & 0.6900 & \textbf{0.7100} & \textbf{0.7101} & 0.7065 & 0.7062 & 0.7051 & 0.7051 & 0.7072 & 0.7075 & 0.7036 & 0.7037 \\
 & PLE-B & 0.7079 & 0.6501 & 0.7016 & 0.4951 & 0.7051 & 0.5654 & \textbf{0.6946} & 0.5520 & 0.7071 & 0.6036 & 0.7055 & 0.4975 & \textbf{0.7104} & 0.5306 & 0.7080 & 0.6234 & 0.7087 & 0.5248 \\
 & PLE-ReLU & 0.7049 & 0.6256 & 0.6973 & 0.4946 & 0.7023 & 0.5702 & 0.6876 & 0.5684 & 0.7041 & 0.6041 & 0.7012 & 0.5115 & 0.7072 & 0.4960 & 0.7035 & 0.6269 & 0.7091 & 0.5225 \\
 & PLR & 0.7063 & 0.6807 & 0.7070 & 0.6824 & 0.7067 & 0.6805 & 0.6861 & 0.6734 & 0.7029 & 0.6799 & 0.7045 & 0.6806 & 0.7093 & 0.6828 & 0.7116 & 0.6866 & 0.7061 & 0.6812 \\
\midrule
\multirow{3}{*}{\shortstack[l]{Adaptive\\coordinates}} & NaryDis-Lite & 0.6920 & 0.7027 & 0.6900 & 0.7007 & 0.6894 & 0.7001 & 0.6764 & 0.6806 & 0.6914 & 0.7032 & 0.6893 & 0.7010 & 0.6891 & 0.7001 & 0.7048 & 0.7105 & 0.6867 & 0.6957 \\
 & NaryDis & 0.7031 & 0.7054 & 0.6996 & 0.7046 & 0.7001 & 0.7043 & 0.6886 & 0.6920 & 0.7016 & 0.7045 & 0.7017 & 0.7042 & 0.7001 & 0.7060 & 0.7089 & 0.7122 & 0.6957 & 0.7052 \\
 & DEER & \textbf{0.7091} & \textbf{0.7088} & \textbf{0.7103} & \textbf{0.7099} & \textbf{0.7099} & \textbf{0.7092} & 0.6924 & 0.6923 & 0.7090 & 0.7092 & \textbf{0.7089} & \textbf{0.7086} & 0.7082 & 0.7077 & \textbf{0.7158} & \textbf{0.7158} & 0.7063 & 0.7066 \\
\midrule
\multirow{3}{*}{\shortstack[l]{Distribution/\\context adaptive}} & DAE & 0.7043 & 0.7005 & 0.7069 & 0.7032 & 0.7074 & 0.7023 & 0.6904 & 0.6887 & 0.7051 & 0.7002 & 0.7045 & 0.6998 & 0.7050 & 0.7004 & 0.7130 & 0.7087 & 0.7034 & 0.6985 \\
 & DAES-Gate & 0.7054 & 0.7053 & 0.7034 & 0.7035 & 0.7043 & 0.7040 & 0.6894 & 0.6895 & 0.7051 & 0.7055 & 0.7053 & 0.7049 & 0.7063 & 0.7064 & 0.7112 & 0.7111 & 0.7097 & \textbf{0.7091} \\
 & DAES-Tran & 0.7046 & 0.7049 & 0.7006 & 0.7002 & 0.7023 & 0.7025 & 0.6899 & 0.6901 & 0.7018 & 0.7020 & 0.7022 & 0.7022 & 0.7031 & 0.7030 & 0.7091 & 0.7090 & 0.7065 & 0.7067 \\
\midrule
\addlinespace[2pt]
\multicolumn{20}{c}{\textbf{AutoML-E}} \\
\midrule
Family & Method & \multicolumn{2}{c}{DNN} & \multicolumn{2}{c}{DeepFM} & \multicolumn{2}{c}{WideDeep} & \multicolumn{2}{c}{NFM} & \multicolumn{2}{c}{PNN} & \multicolumn{2}{c}{DCN} & \multicolumn{2}{c}{DCNv2} & \multicolumn{2}{c}{AutoInt} & \multicolumn{2}{c}{FiBiNET} \\
\cmidrule(lr){3-4}\cmidrule(lr){5-6}\cmidrule(lr){7-8}\cmidrule(lr){9-10}\cmidrule(lr){11-12}\cmidrule(lr){13-14}\cmidrule(lr){15-16}\cmidrule(lr){17-18}\cmidrule(lr){19-20}
 &  & z-score & Raw & z-score & Raw & z-score & Raw & z-score & Raw & z-score & Raw & z-score & Raw & z-score & Raw & z-score & Raw & z-score & Raw \\
\midrule
\multirow{3}{*}{\shortstack[l]{Pointwise\\projection}} & Field & 0.8358 & 0.6322 & \textbf{0.8392} & 0.5000 & \textbf{0.8387} & 0.5000 & 0.8395 & 0.5000 & 0.8384 & 0.5000 & 0.8338 & 0.5000 & 0.8375 & 0.5000 & \textbf{0.8440} & 0.6257 & 0.8341 & 0.5000 \\
 & Linear & 0.8378 & 0.5000 & 0.8378 & 0.4986 & 0.8372 & 0.4997 & 0.8384 & 0.4907 & 0.8383 & 0.5080 & 0.8366 & 0.5016 & 0.8373 & 0.4989 & 0.8388 & 0.5000 & 0.8343 & 0.5035 \\
 & Linear-ReLU & 0.8381 & 0.5000 & 0.8369 & 0.5000 & 0.8373 & 0.5001 & 0.8367 & 0.5039 & 0.8391 & 0.4969 & 0.8362 & 0.5038 & 0.8373 & 0.5068 & 0.8398 & 0.5000 & 0.8344 & 0.4966 \\
\midrule
\multirow{3}{*}{\shortstack[l]{Discretization}} & Quantile & \textbf{0.8387} & \textbf{0.8387} & 0.8383 & 0.8383 & 0.8375 & 0.8375 & 0.8396 & 0.8396 & 0.8386 & 0.8386 & 0.8350 & 0.8350 & 0.8357 & 0.8357 & 0.8424 & 0.8424 & 0.8337 & 0.8337 \\
 & LogBucket & 0.8366 & 0.8382 & 0.8372 & 0.8365 & 0.8374 & 0.8362 & 0.8385 & 0.8389 & 0.8385 & 0.8371 & 0.8352 & 0.8357 & 0.8363 & 0.8347 & 0.8414 & 0.8410 & 0.8325 & 0.8338 \\
 & AutoDis & 0.8364 & 0.8351 & 0.8375 & 0.8370 & 0.8378 & 0.8368 & 0.8397 & 0.8386 & 0.8383 & 0.8371 & 0.8367 & 0.8356 & 0.8370 & 0.8354 & 0.8431 & 0.8415 & 0.8331 & 0.8325 \\
\midrule
\multirow{2}{*}{\shortstack[l]{Fixed basis}} & Periodic & 0.8379 & 0.8370 & 0.8375 & 0.8376 & 0.8373 & \textbf{0.8384} & 0.8375 & 0.8376 & 0.8379 & 0.8390 & \textbf{0.8380} & 0.8380 & 0.8379 & 0.8384 & 0.8395 & 0.8371 & 0.8342 & \textbf{0.8388} \\
 & B-spline & 0.8373 & 0.8373 & 0.8376 & 0.8374 & 0.8372 & 0.8374 & 0.8397 & 0.8398 & 0.8391 & 0.8387 & 0.8371 & 0.8370 & 0.8373 & 0.8373 & 0.8417 & 0.8420 & 0.8332 & 0.8332 \\
\midrule
\multirow{4}{*}{\shortstack[l]{Piecewise basis}} & PLE-A & 0.8375 & 0.8377 & 0.8385 & 0.8385 & 0.8379 & 0.8376 & 0.8375 & 0.8375 & 0.8395 & \textbf{0.8396} & 0.8378 & \textbf{0.8382} & \textbf{0.8393} & \textbf{0.8394} & 0.8395 & 0.8393 & \textbf{0.8364} & 0.8364 \\
 & PLE-B & 0.8366 & 0.5000 & 0.8377 & 0.4986 & 0.8378 & 0.5001 & 0.8378 & 0.4973 & \textbf{0.8396} & 0.5046 & 0.8378 & 0.4970 & 0.8376 & 0.4976 & 0.8411 & 0.5000 & 0.8353 & 0.4971 \\
 & PLE-ReLU & 0.8367 & 0.5000 & 0.8383 & 0.4986 & 0.8369 & 0.5001 & 0.8378 & 0.4973 & 0.8385 & 0.5046 & 0.8374 & 0.4970 & 0.8376 & 0.4976 & 0.8401 & 0.5000 & 0.8362 & 0.4971 \\
 & PLR & 0.8381 & 0.8387 & 0.8388 & 0.8379 & 0.8370 & 0.8374 & 0.8382 & \textbf{0.8403} & 0.8389 & 0.8383 & 0.8370 & 0.8380 & 0.8375 & 0.8370 & 0.8405 & 0.8383 & 0.8323 & 0.8386 \\
\midrule
\multirow{3}{*}{\shortstack[l]{Adaptive\\coordinates}} & NaryDis-Lite & 0.8369 & 0.8377 & 0.8377 & 0.8379 & 0.8376 & 0.8376 & 0.8380 & 0.8382 & 0.8382 & 0.8384 & 0.8357 & 0.8351 & 0.8363 & 0.8364 & 0.8418 & 0.8419 & 0.8329 & 0.8336 \\
 & NaryDis & 0.8371 & 0.8367 & 0.8366 & 0.8371 & 0.8365 & 0.8366 & 0.8393 & 0.8394 & 0.8380 & 0.8386 & 0.8358 & 0.8352 & 0.8368 & 0.8369 & 0.8417 & \textbf{0.8431} & 0.8312 & 0.8321 \\
 & DEER & 0.8374 & 0.8371 & 0.8390 & \textbf{0.8391} & 0.8385 & 0.8382 & \textbf{0.8400} & 0.8399 & 0.8379 & 0.8382 & 0.8367 & 0.8364 & 0.8375 & 0.8377 & 0.8424 & 0.8425 & 0.8352 & 0.8351 \\
\midrule
\multirow{3}{*}{\shortstack[l]{Distribution/\\context adaptive}} & DAE & 0.8384 & 0.8379 & 0.8383 & 0.8381 & 0.8380 & 0.8371 & 0.8393 & 0.8392 & 0.8389 & 0.8384 & 0.8368 & 0.8365 & 0.8363 & 0.8361 & 0.8425 & 0.8420 & 0.8344 & 0.8344 \\
 & DAES-Gate & 0.8383 & 0.8384 & 0.8359 & 0.8357 & 0.8357 & 0.8357 & 0.8388 & 0.8389 & 0.8389 & 0.8387 & 0.8349 & 0.8351 & 0.8378 & 0.8378 & 0.8421 & 0.8424 & 0.8316 & 0.8317 \\
 & DAES-Tran & 0.8382 & 0.8384 & 0.8373 & 0.8373 & 0.8371 & 0.8376 & 0.8388 & 0.8389 & 0.8378 & 0.8383 & 0.8371 & 0.8371 & 0.8361 & 0.8363 & 0.8414 & 0.8414 & 0.8334 & 0.8335 \\
\bottomrule
\end{tabular}%
}
\end{table*}